\documentclass[stslayout, noinfoline]{imsart}
\usepackage{geometry} 
\usepackage{graphicx}	
\usepackage{amssymb, amsmath}
\usepackage{epstopdf}
\usepackage{natbib}
\usepackage{subfigure}
\usepackage{algcompatible}
\usepackage[floatrow]{trivfloat}
\usepackage{comment}
\usepackage[colorlinks,citecolor=blue,urlcolor=blue]{hyperref}

\newcommand\dd{\mathrm{d}}
\newcommand\X{\mathcal{X}}
\newcommand\E{\mathcal{E}}
\newcommand\R{\mathbb{R}}

\graphicspath{{figures/}}

\begin{document} 

\begin{frontmatter}

\title{On Symplectic Optimization}
\runtitle{On Symplectic Optimization}

\begin{aug}
  \author{Michael Betancourt%
  \ead[label=e1]{betanalpha@gmail.com}}

  \author{Michael I. Jordan%
  \ead[label=e2]{jordan@cs.berkeley.edu}}
  
  \author{Ashia C. Wilson}

  \runauthor{Betancourt Et Al.}

  \address{Michael Betancourt is a research scientist at Symplectomorphic, LLC \printead{e1}.          
           Michael I. Jordan is the Pehong Chen Distinguished Professor in the Departments 
           of EECS and Statistics at the University of California, Berkeley \printead{e2}.
           Ashia C. Wilson is a graduate student in the Department of Statistics at the
           University of California, Berkeley. }

\end{aug}

\begin{abstract}
Accelerated gradient methods have had significant impact
in machine learning---in particular the theoretical side of 
machine learning---due to their ability to achieve oracle 
lower bounds.  But their heuristic construction has hindered
their full integration into the practical machine-learning 
algorithmic toolbox, and has limited their scope.  In this 
paper we build on recent work which casts acceleration as 
a phenomenon best explained in continuous time, and we augment 
that picture by providing a systematic methodology for converting 
continuous-time dynamics into discrete-time algorithms while
retaining oracle rates.  Our framework is based on ideas from 
Hamiltonian dynamical systems and symplectic integration.
These ideas have had major impact in many areas in applied
mathematics, but have not yet been seen to have a relationship 
with optimization.
\end{abstract}

\end{frontmatter}

\section{Introduction}

Optimization theory has played an increasingly central role in the 
development of machine learning in recent years.  This has happened 
not only because optimization theory supplies algorithms and convergence 
rates for learning algorithms, but also because it supplies lower bounds, 
and hence fundamental understanding.  A milestone in this regard 
was the discovery by \citet{nemirovskii1983problem} of oracle 
lower bounds for gradient-based optimization, and the ensuing 
derivation by \citet{nesterov1983method} of an ``accelerated gradient 
descent'' (AGD) algorithm whose rate is provably better than that 
of gradient descent, and which matches the oracle lower bound.

A flurry of mathematical and algorithmic results have followed
in the wake of these seminal discoveries from the 1980's, but even after
three decades there remains a lack of understanding of the general 
acceleration phenomenon.  In particular, a theoretical framework 
that can \emph{generate} accelerated methods has not yet emerged.
Recent progress in this regard has been achieved by considering 
continuous-time analogs of acceleration methods~\citep{su2016differential,
Krichene15}.  Notably, \citet{WibisonoEtAl:2016} presented a variational
framework, involving a ``Bregman Lagrangian,'' that generates
differential equations with rates that are continuous-time
analogs of the discrete-time oracle rates.

The work of \citet{WibisonoEtAl:2016}, however, only partially addresses 
the problem of providing a generative framework for acceleration.
They show that any desired rate can be achieved in continuous 
time---different algorithms follow the same \emph{path} in phase 
space while doing so at different \emph{speeds}, and that the speed 
can be arbitrarily fast.  Differences in speed thus correspond to a 
mere change of the clock by which time is measured.  The fact that 
lower bounds for rates emerge in discrete time must therefore have 
something to do with the discretization of the class of differential 
equations arising from the Bregman Lagrangian.  \citet{WibisonoEtAl:2016} 
were able to provide an adhoc discretization that yielded an algorithm 
whose rate matches the rate of Nesterov acceleration in a particular setting, 
but their framework is silent on a general methodology for providing 
such discretizations.

The class of differential equations arising from the Bregman 
Lagrangian must be special in some sense, given that they deliver 
continuous-time analogs of oracle rates.  The notion that certain 
differential equations are special has a long history in physics, 
where underlying Lagrangians and Hamiltonians possess certain 
mathematical symmetries that yield conservation laws for the 
resulting differential equations.  Moreover, the 
venerable field of symplectic integration shows that it is 
possible to preserve these conservation laws when discretizing
the differential equations~\citep{HairerEtAl:2006}.  The resulting integrators 
improve upon classical integrators, e.g., Euler, Runge-Kutta, precisely 
because they respect the underlying mathematical symmetries.  This 
results in certain error terms canceling, with a variety of favorable 
consequences, including long-term stability.  Of particular relevance 
to the current setting, the stability of such integrators means that 
it is possible to take much larger step sizes than with classical 
integrators.  Given that we are interested in ``accelerated'' methods 
that arrive at an optimum as quickly as possibly, this feature of 
symplectic integration seems directly relevant.

In the current paper we show how to apply symplectic integration to
gradient-based optimization.  Our approach is cast in a Hamiltonian 
framework, obtained from the Bregman-Lagrangian framework via a Legendre
transformation.  This Hamiltonian is time-varying, a fact that we
address via a lifting procedure.  We then show how to derive a 
symplectic integrator from the lifted Hamiltonian.  The end result 
is a fully generative mathematical pipeline, from problem specification 
to discrete-time accelerated algorithm.  Our algorithms 
are related to Nesterov's algorithms, but they are \emph{not} exactly 
the same, and the differences are interesting.

\section{Bregman Dynamics for Optimization}
\label{sec:dynamics}

We begin with a brief review of the dynamical framework 
introduced in \citet{WibisonoEtAl:2016}, including the heuristic 
discretization that those authors employed to obtain accelerated 
discrete-time optimization algorithms.  

Consider a Euclidean vector space, $\X$, which we will denote as 
the \emph{configuration manifold}.  This configuration manifold 
is equipped with the Euclidean gradient operator, $\nabla$, 
and inner product, $\left<\, ,\right>$.  Formally we should
be careful to distinguish between points, vectors, and covectors 
on the configuration manifold but we will reserve that level of 
rigor for a more formal geometric treatment presented in
Appendix~\ref{apx:geometric}.

Given a smooth \emph{objective function} on the configuration
manifold, $f : \X \rightarrow \R$, the optimization problem
is then to compute minima of $f$:
\begin{equation*}
x^{*} \in \underset{x \in \X}{\mathrm{argmin}} f(x).
\end{equation*}
In accordance with most of the theoretical literature on
acceleration, we will focus on the setting in which $f$ is 
convex and exhibits a single minimum.  But it is worth 
emphasizing that convexity is not needed in our construction of 
the Hamiltonian nor for the symplectic integrators.  Note, 
moreover, that there is a growing literature on acceleration 
in the non-convex setting, where the dynamical systems 
perspective presented here is also proving to be useful; 
see, e.g., \citet{JinNetrapalliJordan}.

From a dynamical perspective, the objective function naturally
plays the role of a potential energy, with the minimum at 
the basin of that potential.  If we want to generate dynamics
that might settle into this basin, however, then we need to 
consider not the configuration manifold but rather its 
\emph{tangent bundle}, $T \X$, consisting of points in $\X$ 
paired with tangent vectors or \emph{velocities}. In particular, 
we need to complement the potential energy with a kinetic energy 
function on the tangent bundle.

Following \citet{WibisonoEtAl:2016} we construct a kinetic 
energy from an auxiliary smooth function, $h : \X \rightarrow \R$, 
and its associated \emph{Bregman divergence},
$D_{h} (y, x) = h(y) - h(x) - \left< \nabla h (x), y - x \right>$.
For any given point in the tangent bundle, $(x, v) \in T \X$, 
we can translate the base point, $x$, in the direction of the 
velocity, $v$, to give the new point $x' = x + e^{-\alpha(t)} v$,
for a scaling function $\alpha(t)$.  The divergence between these 
two points defines the \emph{Bregman kinetic energy}:
\begin{align*}
K(x, v) 
&\equiv D_{h} (x + e^{-\alpha(t)} v, x)
\\
&= h(x + e^{-\alpha(t)} v) - h(x) 
- e^{-\alpha(t)} \left< \nabla h (x), v \right>.
\end{align*}
This kinetic energy admits the evocative interpretation as a 
comparison of how $h$ changes under a finite translation:
\begin{equation*}
\Delta h(x, v, t) = h(x + e^{-\alpha(t)} v) - h(x),
\end{equation*}
versus a scaled infinitesimal translation:
\begin{equation*}
\delta h(x, v, t) = e^{-\alpha(t)} \left< \nabla h (x), v \right>.
\end{equation*}
Defining a time-dependent potential energy,
\begin{equation*}
U(x, t) = e^{\beta(t)} f(x),
\end{equation*}
we can then construct the \emph{Bregman Lagrangian} as:
\begin{align*}
L(x, v, t) 
&= e^{\alpha(t) + \gamma(t)} (K(x, v) - U(x))
\\
&= e^{\alpha(t) + \gamma(t)} 
\left(  D_{h} (x + e^{-\alpha(t)} v, x) - e^{\beta(t)} f(x) \right).
\end{align*}

From the Bregman Lagrangian we obtain a variational problem on the 
tangent bundle whose solutions yield smooth trajectories via 
the ordinary differential equations
\begin{equation*}
\frac{\dd}{\dd t} \frac{\partial L}{ \partial v} (x, v, t)
= \frac{\partial L}{\partial x} (x, v, t).
\end{equation*}
The time dependence of the Lagrangian allows the dynamics to 
rapidly converge to a minimum, as opposed to dynamics obtained
from a time-independent Lagrangian which would oscillate around 
the desired minimum. 

\citet{WibisonoEtAl:2016} also defined the following 
\emph{ideal scaling conditions}:
\begin{align*}
\alpha(t) &= \log p - \log t
\\
\beta(t) &= p \log t + \log C
\\
\gamma(t) &= p \log t,
\end{align*}
for $p, C \in \R^{+}$, and demonstrated that if $f$ and $h$ are
sufficiently well behaved then the Bregman dynamics will provably
converge to the minimum of $f$ at the polynomial rate, 
$\mathcal{O} (1 / t^{p})$.  This captures not only classical 
Nesterov acceleration, with its rate of $\mathcal{O} (1 / t^{2})$, 
but also higher-order accelerated algorithms for which $p > 2$.

Unfortunately it is not obvious how to discretize these continuous
dynamics to obtain a discrete-time algorithm.  \citet{WibisonoEtAl:2016} 
found that simple discretizations yield algorithms that do not not 
recover accelerated Nesterov methods and can even be unstable.  
Ultimately they were able to find a stable discretization that 
yielded the oracle rates.  Their discretization was a sophisticated 
but heuristic discretization, coupling a Crank-Nicolson discretization 
of the position updates and a backwards Euler discretization of the 
velocity updates with a third implicit sequence of intermediate positions, 
$(y_n)$:
\begin{align*}
x_{n + 1} &= \frac{p}{n + p} z_{n} + \frac{n}{n + p} y_{n}
\\
y_{n} &= \underset{y \in \X}{\mathrm{argmin}}
\left[ f_{p - 1}(y, ; x_{n}) 
+ \frac{N}{\epsilon^{p} \, p} || y - x_{n} ||^{p} \right]
\\
z_{n} &= \underset{z \in \X}{\mathrm{argmin}}
\left[ C \, p \, n^{p - 1} \left< \nabla f(y_{n}), z \right>
+ \frac{1}{\epsilon^{p}} D_{h} (z, z_{n - 1}) \right].
\end{align*}
Here $f_{p - 1}(y; x_{n})$ is the order-$p$ Taylor
expansion of the objective function around $x_{n}$.
This third sequence proved to be the key, stabilizing
the discretization and preserving the convergence rates 
of the continuous-time dynamics.  This discretization was
obtained by analogy with Nesterov's classical updates, and 
in this paper we will refer to these discretizations 
as \emph{generalized Nesterov discretizations}.

\section{Simulating Bregman Dynamics with Symplectic Integrators}

The difficulties associated with discretizing Lagrangian dynamics 
are well known in the mathematical literature~\citep{LeimkuhlerEtAl:2004, 
HairerEtAl:2006}.  Discretizations of Lagrangian dynamics 
are often fragile, especially when the dimension of the configuration 
space is large.  Even high-order discretizations can diverge after short 
integration times.  Ultimately this is because any discretization of 
dynamics on the tangent bundle does not preserve the continuous symmetries 
of the dynamical system that stabilize the exact dynamics.

Fortunately we can readily construct a discretization that 
\emph{does} preserve the necessary symmetries, by exploiting the 
dual \emph{Hamiltonian} representation of the Bregman 
dynamics.  The Hamiltonian system is the Legendre transform
of the Lagrangian system, trading velocities, $v$, and the
tangent bundle, $T\X$, for dual \emph{momenta}, $r$, and 
the \emph{cotangent bundle}.

In this section we will first construct the Hamiltonian
representation of the Bregman dynamics and then modify
that system to circumvent the explicit time dependence and
admit the application of symplectic integrators.

\subsection{The Bregman Hamiltonian}

To build up the Bregman Hamiltonian from the Bregman 
Lagrangian we first have to relate define momenta as
the derivative of the Lagrangian with respect to the
velocities,
\begin{align*}
r (x, v, t)
&= \frac{\partial L}{ \partial v} (x, v, t) \\
&= e^{\gamma(t)} 
\left( \frac{\partial h}{\partial x} (x + e^{-\alpha(t)} v)
- \frac{\partial h}{\partial x}(x) \right).
\end{align*}
Given the Legendre conjugate of $h$,
\begin{align*}
h^{*} = \underset{v \in T\X}{\mathrm{sup}} \left[ r \cdot v - h(v) \right],
\end{align*}
we can invert this relationship to give
\begin{equation*}
v(x, r, t) = e^{\alpha(t)} \left( 
\frac{\partial h^{*}}{\partial r} (e^{-\gamma(t)} r 
+ \frac{\partial h}{\partial x} (x) ) - r \right).
\end{equation*}
We are now in position to construct the \emph{Bregman Hamiltonian}:
\begin{align*}
H(x, r, t) 
&= r \cdot v(x, r, t) - L(x, v(r), t)
\\
&=
e^{\alpha(t) + \gamma(t)} \left(
D_{h^{*}}(e^{-\gamma(t)} r + \frac{\partial h}{\partial x} (x), \right.
\left. \frac{ \partial h }{ \partial x} (x) )
+ e^{\beta(t)} f(x) \right)
\\
&=
e^{\alpha(t) + \gamma(t)} \left(
D_{h^{*}}(e^{-\gamma(t)} r + \nabla h (x), \nabla h (x) )
+ e^{\beta(t)} f(x) \right),
\end{align*}
where
\begin{equation*}
D_{h^{*}} (r, s) = h^{*} (r) - h^{*}(s) - 
\frac{\partial h^{*}}{\partial r} (s) \cdot (p - s).
\end{equation*}

The Bregman dynamics can then be generated from the Bregman
Hamiltonian by integrating \emph{Hamilton's equations}:
\begin{equation*}
\frac{\dd x}{\dd t} 
=
+ \frac{\partial H}{\partial r} (x, r, t), \hspace{1cm}
\frac{\dd r}{\dd t}
=
- \frac{\partial H}{\partial x} (x, r, t).
\end{equation*}
When the Hamiltonian does not explicitly depend on time the 
dynamics are said to be \emph{autonomous} \citep{JoseEtAl:1998} 
and there are standard methods for constructing symplectic
integrators that preserve the critical symmetries that stabilize 
the dynamics.  These integrators are extremely accurate, 
defining discretized dynamics that mirror the exact dynamics 
even for long integration times and high-dimensional 
configuration spaces \citep{LeimkuhlerEtAl:2004, HairerEtAl:2006}.

Unfortunately the explicit time dependence that allows the 
dynamics to converge to the minimum of the objective also 
renders the Bregman Hamiltonian \emph{non-autonomous}.
Fortunately this problem can be circumvented.  As we show 
in the next section, by introducing a few more auxiliary 
variables we can lift the non-autonomous Bregman Hamiltonian 
system into an autonomous, \emph{extended} Hamiltonian system 
where symplectic integrators are immediately applicable.

\subsection{Making the non-autonomous autonomous}

For an autonomous dynamical system time is simply a 
parameterization of motion along a dynamical trajectory.
In particular, we can utilize uniform time increments 
to facilitate stable discretization of those dynamics.  
When the trajectories themselves explicitly depend on 
time, however, stable discretizations become all the 
more challenging.  In order to overcome this difficulty 
we need to decouple the two responsibilities of time in 
a dynamical system by incorporating the explicit time 
into the configuration space and introducing a new 
effective time to parameterize motion along the dynamical 
trajectories \citep{deLeonEtAl:1989}.

The explicit time, $t$, serves as a new position in the 
extended configuration space, $(x, t) \in \Xi$. The 
extended cotangent bundle then includes a conjugate 
energy, $(x, t, r, \E) \in T^{*} \Xi$.  We then define 
the extended Hamiltonian as
\begin{align*}
H_{\Xi} = \E - H(x, t, r).
\end{align*}
Here the conjugate energy $\E$ must compensate 
for the time dependence of the original Hamiltonian
to ensure that the extended Hamiltonian, $H_{\Xi}$, 
is constant along dynamical trajectories.

The corresponding equations of motion for the extended
Hamiltonian system become
\begin{align*}
\frac{\dd x}{\dd \tau} 
&=
+ \frac{\partial H_{\Xi}}{\partial r} (x, t, r, \E)
\\
\frac{\dd t}{\dd \tau} 
&=
+ \frac{\partial H_{\Xi}}{\partial \E} (x, t, r, \E)
\\
\frac{\dd r}{\dd \tau}
&=
- \frac{\partial H_{\Xi}}{\partial x} (x, t, r, \E)
\\
\frac{\dd \E}{\dd \tau}
&=
- \frac{\partial H_{\Xi}}{\partial t} (x, t, r, \E),
\end{align*}
which introduces the effective time, $\tau$, to parameterize 
the motion along the extended dynamical trajectories.  These 
dynamics projected back down to the original cotangent bundle 
yield the original Bregman dynamics, but by separating the 
time dependence of the dynamics from the parameterization of 
the trajectories these extended dynamics are manifestly autonomous.  
In particular, we can immediately apply symplectic integrators 
to the extended Hamiltonian system.  

\subsection{Building an extended leapfrog integrator}
\label{sec:leapfrog}

We are now in a position to build a symplectic integrator 
to simulate the Bregman dynamics.  There is a rich literature 
on symplectic integrators and their exceptional performance
\citep{LeimkuhlerEtAl:2004, HairerEtAl:2006}, so here we 
will limit our discussion to the construction and behavior 
of a simple \emph{leapfrog integrator} for our extended 
Hamiltonian system.  Despite the simplicity of this integrator,
we will see in Section \ref{sec:experiments} that it rivals 
the performance of the generalized Nesterov discretization.

Symplectic integrators are naturally constructed by splitting 
the Hamiltonian into component Hamiltonians whose dynamics 
can be solved exactly, or at least sufficiently close to 
exactly numerically, and then composing those dynamics 
together symmetrically.  For example, consider the splitting
\begin{equation*}
H_{\Xi} = H_{A} + H_{B} + H_{C},
\end{equation*}
where
\begin{align*}
H_{A} (\E) &= \E
\\
H_{B} (x, r, t) &= e^{\alpha(t) + \gamma(t)}
D_{h^{*}}(e^{-\gamma(t)} r + \frac{\partial h}{\partial x} (r),
\frac{ \partial h }{ \partial x} (x) )
\\
H_{C} (x, t) &= e^{\alpha(t) + \gamma(t) + \beta(t)} f(x).
\end{align*}
These three component Hamiltonians generate dynamics in the 
extended cotangent bundle with the six vector fields,
\begin{align*}
\vec{H}_{A} &=
+ \frac{ \partial H_{1} }{ \partial \E }
\! \left( \E \right) \frac{ \dd }{ \dd t}
\\
\vec{H}_{B1} &=
- \frac{ \partial H_{2} }{ \partial x }
\! \left(x, r, t \right) \frac{ \dd}{ \dd r}
\\
\vec{H}_{B2} &=
- \frac{ \partial H_{2} }{ \partial t }
\! \left(x, r, t \right) \frac{ \dd }{ \dd \E}
\\
\vec{H}_{B3} &=
+ \frac{ \partial H_{2} }{ \partial r }
\! \left(x, r, t \right) \frac{ \dd }{ \dd x} 
\\
\vec{H}_{C1} &=
- \frac{ \partial H_{3} }{ \partial x }
\! \left(x, t \right) \frac{ \dd }{ \dd r } 
\\
\vec{H}_{C2} &=
-\frac{ \partial H_{3} }{ \partial t }
\! \left(x, t \right) \frac{ \dd }{ \dd \E}.
\end{align*}

Regardless of the nature of $h$, the vector fields 
$\vec{H}_{A}$, $\vec{H}_{B2}$, $\vec{H}_{C1}$, and $\vec{H}_{C2}$ 
will always be trivial and hence their evolution can be 
solved exactly.  For example,
\begin{equation*}
\exp \left( \epsilon \vec{H}_{C1} \right) \big( x, t, r, \E \big)
=
\big( x, t, r - \epsilon \frac{\partial H_{3}}{\partial x}(x, t), \E \big).
\end{equation*}
On the other hand the component dynamics of $\vec{H}_{B1}$
and $\vec{H}_{B3}$ may or may not be trivial, depending on 
the choice of $h$.  Even if they are nonlinear, however, they 
can be solved implicitly using, for example, fixed-point 
iterations.

We can now build a symplectic integrator by composing these 
component dynamics together to approximate the full dynamics.  
Here we will consider a symmetric \emph{leapfrog} composition,
\begin{align*}
\Phi_{\epsilon}
&= \;\;
\exp \left( \frac{\epsilon}{2} \vec{H}_{A} \right)
\, \circ \,
\exp \left( \frac{\epsilon}{2} \left( \vec{H}_{B2} + \vec{H}_{C2} \right) \right)
\circ
\exp \left( \frac{\epsilon}{2} \vec{H}_{C1} \right) \\
& \quad
\circ
\exp \left( \frac{\epsilon}{2} \vec{H}_{B1} \right)
\circ
\exp \left( \epsilon \vec{H}_{B3} \right)
\circ
\exp \left( \frac{\epsilon}{2} \vec{H}_{B1} \right) \\
& \quad
\circ
\exp \left( \frac{\epsilon}{2} \vec{H}_{C1} \right)
\circ
\exp \left( \frac{\epsilon}{2} \left( \vec{H}_{B2} + \vec{H}_{C2} \right) \right)
\circ
\exp \left( \frac{\epsilon}{2} \vec{H}_{A} \right).
\end{align*}
Applying the Baker-Campell-Hausdorff equation to this
composite operator demonstrates that the symmetry of 
the composition ensures the cancellation of all terms 
linear and quadratic in the step size, leaving
\begin{equation*}
\Phi_{\epsilon} = \exp \left( \epsilon \vec{H} \right) 
+ \mathcal{O}(\epsilon^{3}).
\end{equation*}
If we apply the composite operator for $N = T / \epsilon$ 
steps we then we can approximate the evolution of the exact 
dynamics for time $T$ with error only quadratic in the step 
size:
\begin{equation*}
\left( \Phi_{\epsilon} \right)^{N}
= 
\left( \exp \left( \epsilon \vec{H} \right) \right)^{N}
+ \frac{T}{\epsilon} \mathcal{O}(\epsilon^{3})
=
\exp \left( T \vec{H} \right)
+ \mathcal{O}(\epsilon^{2}).
\end{equation*}

Because each component operator exactly solves a dynamical
system, their solutions preserve the dynamical symmetries 
that maintain stable evolution.  Moreover, the symmetric 
composition of these component dynamics yields an approximate 
dynamics that very accurately tracks the dynamics of the 
extended Hamiltonian system even for long integration times, 
at least if the step size is small enough that the expansion 
converges.  Although the approximate dynamics of a symplectic 
integrator will diverge if the step size is too large, the 
inherent stability of this approximation admits much larger
step sizes, and hence reduced computation, than other 
discretizations of the dynamics.

This symmetric leapfrog integrator enjoys a global error
quadratic in the step size and consequently it is classified
as a second-order integrator.  Higher-order integrators can
just as easily be built up by applying each component operator 
multiple times in careful arrangements to cancel more and more 
error terms.

\section{Experiments}
\label{sec:experiments}

To explore the performance of symplectic optimization numerically, 
we consider a relatively simple experiment.  Let $\X$ be a 
50-dimensional Euclidean space equipped with the quadratic 
objective function
\begin{equation*}
f(x) = \left<\Sigma^{-1} x, x \right>,
\end{equation*}
where
\begin{equation*}
\Sigma_{ij} = \rho^{|i - j|},
\end{equation*}
and $\rho = 0.9$ to correlate the objective.

Let the auxiliary function, $h$, also be quadratic but 
without any interactions among the coordinates,
\begin{equation*}
h(x) = \left<x, x \right>.
\end{equation*}
The Bregman Hamiltonian becomes
\begin{equation*}
H(x, r, t) = 
\frac{1}{2} e^{\alpha(t) - \gamma(t)} \left<p, p \right>
+ e^{\alpha(t) + \beta(t) + \gamma(t)} f \! \left( x \right),
\end{equation*}
and the component vector fields of the extended dynamics
take the form
\begin{align*}
\vec{H}_{A} &= \frac{ \dd }{ \dd t}
\\
\vec{H}_{B1} &= 0
\\
\vec{H}_{B2} &= 
- \frac{1}{2} \left(
\frac{ \partial \alpha }{ \partial t }(t) 
 - \frac{ \partial \gamma }{ \partial t }(t) \right) 
e^{\alpha(t) - \gamma(t)} \left<p, p\right>
\frac{ \dd }{ \dd \E }
\\
\vec{H}_{B3} &=
e^{\alpha(t) - \gamma(t)} r \frac{ \dd }{ \dd x}
\\
\vec{H}_{C1} &=
- e^{\alpha(t) + \beta(t) + \gamma(t)} \nabla f \! \left( x \right)
\frac{ \dd }{ \dd r }
\\
\vec{X}_{C2} &=
- \left( \frac{ \partial \alpha }{ \partial t }(t)
+ \frac{ \partial \beta }{ \partial t }(t)
+  \frac{ \partial \gamma }{ \partial t }(t) \right) 
e^{\alpha(t) + \beta(t) + \gamma(t)} f \! \left( x \right)
\frac{ \dd }{ \dd \E }.
\end{align*}
In this case each of these vector fields are trivial and
hence can be integrated exactly.

Finally we adopt the ideal scaling conditions for $\alpha(t), 
\beta(t)$ and $\gamma(t)$ discussed in Section \ref{sec:dynamics},
in which case the vector fields become
\begin{align*}
\vec{H}_{A} &= \frac{ \dd}{ \dd t} 
\\
\vec{H}_{B1} &= 0
\\
\vec{H}_{B2} &=
\frac{1}{2} \frac{p (p + 1)}{t^{p + 2}} \left<r, r\right>
\frac{ \dd }{ \dd \E }
\\
\vec{H}_{B3} &=
\frac{p}{t^{p + 1}} r
\frac{ \dd }{ \dd x }
\\
\vec{H}_{C1} &=
- C \, p \, t^{2p - 1} \nabla f \! \left( x \right)
\frac{ \dd }{ \dd r }
\\
\vec{H}_{C2} &=
- C \, p (2p - 1) \, t^{2p - 2} f \! \left( x \right)
\frac{ \dd }{ \dd \E }.
\end{align*}
Applying these component dynamics to the second-order leapfrog 
integrator introduced in Section \ref{sec:leapfrog} then gives
the symmetric update sequence
\begin{align*}
t_{n + \frac{1}{2}} &= t_{n} + \epsilon
\\
\E_{n + \frac{1}{2}} &= \E_{n} + \epsilon
\left( \frac{1}{2} \frac{p (p + 1)}{t^{p + 2}} \left< r_{n}, r_{n} \right>
+ - C \, p (2p - 1) \, t^{2p - 2} f \! \left( x_{n} \right) \right)
\\
r_{n + \frac{1}{2}} &= r_{n}
- \epsilon \, C \, p \, t^{2p - 1} \nabla f \! \left( x_{n} \right)
\\
x_{n + 1} &= x_{n} 
+ \epsilon \, \frac{p}{t^{p + 1}} r_{n + \frac{1}{2}}
\\
r_{n + 1} &= r_{n + \frac{1}{2}}
- \epsilon \, C \, p \, t^{2p - 1} \nabla f \! \left( x_{n + 1} \right)
\\
\E_{n + 1} &= \E_{n + \frac{1}{2}} + \epsilon
\left( \frac{1}{2} \frac{p (p + 1)}{t^{p + 2}} \left< r_{n + 1}, r_{n + 1} \right>
+ - C \, p (2p - 1) \, t^{2p - 2} f \! \left( x_{n + 1} \right) \right)
\\
t_{n + 1} &= t_{n + \frac{1}{2}} + \epsilon.
\end{align*}

For comparison we implement the three-step dynamical
Nesterov discretization derived in \citet{WibisonoEtAl:2016}. 
Given the ideal scaling conditions and the quadratic
auxiliary function, $h$, this algorithm is given by
\begin{align*}
x_{n + 1} &= 
\frac{p}{n + 1} z_{n} 
+ (1 - \frac{p}{n + 1} ) y_{n}
\\
y_{n + 1} &= x_{n + 1} 
- \frac{p \, \epsilon^{p} }{2 N} 
\nabla f(x_{n + 1})
\\
z_{n + 1} &= z_{n} 
- \epsilon^{p} \, C \, p \, (n + 1)^{p - 1} 
\nabla f( y_{n + 1} ).
\end{align*}
For both the extended Hamiltonian system and the Nesterov
sequence we take $\epsilon = 0.1$, $p = 2$, $C = 0.0625$,
and $N = 2$.  

\begin{figure*}
\centering
\subfigure[]{ \includegraphics[width=2.8in]{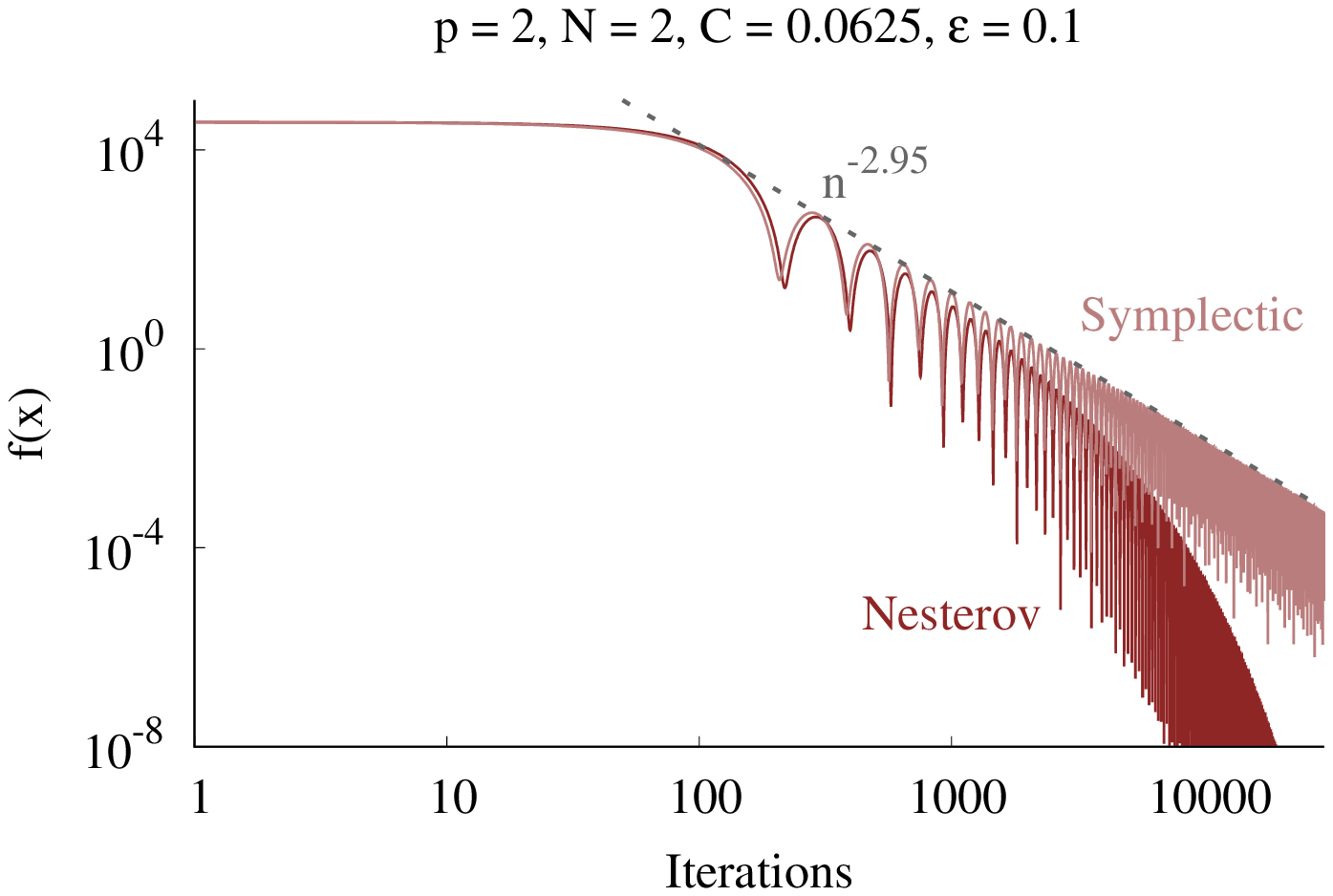} }
\subfigure[]{ \includegraphics[width=2.8in]{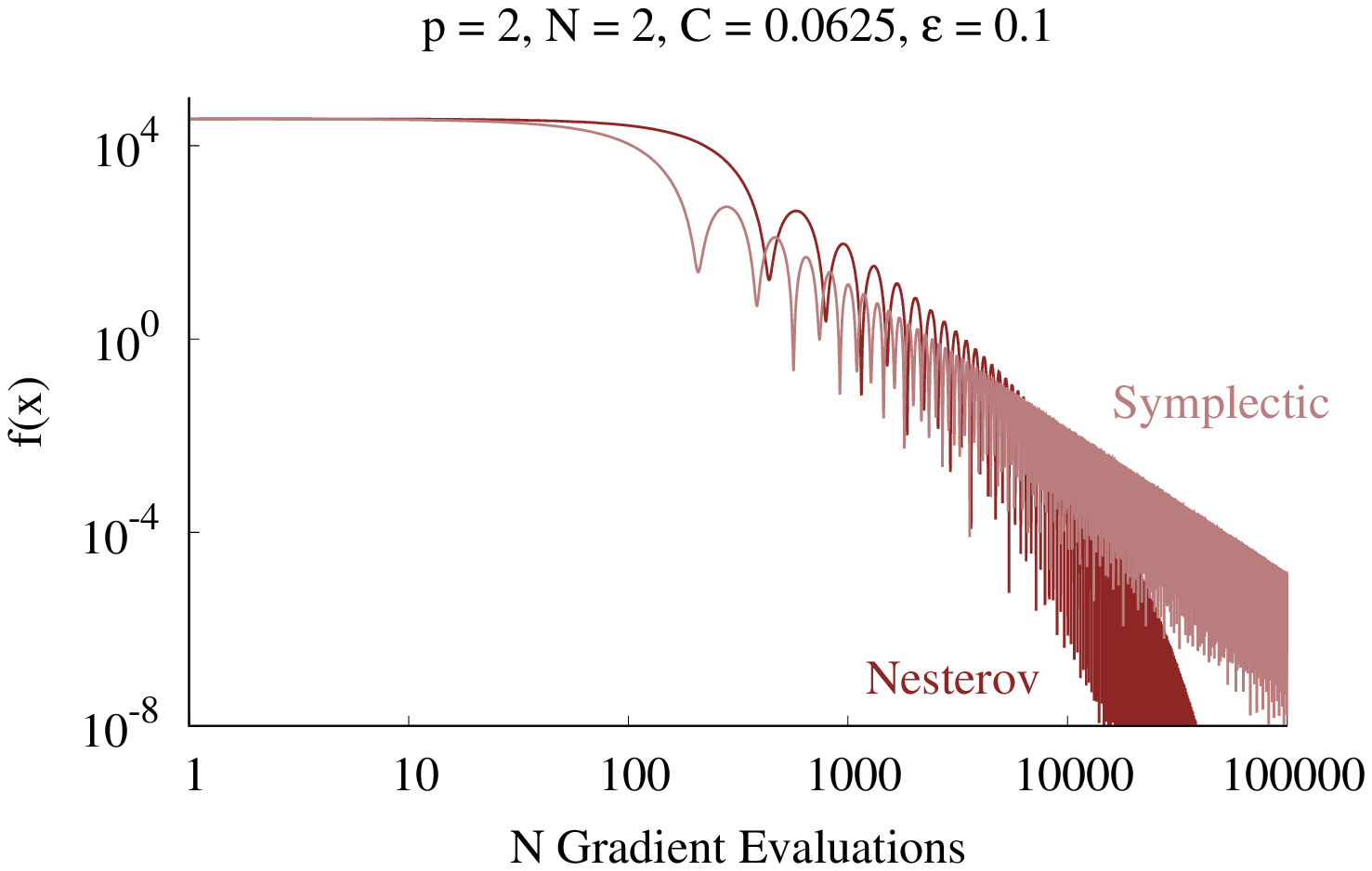} }
\subfigure[]{ \includegraphics[width=2.8in]{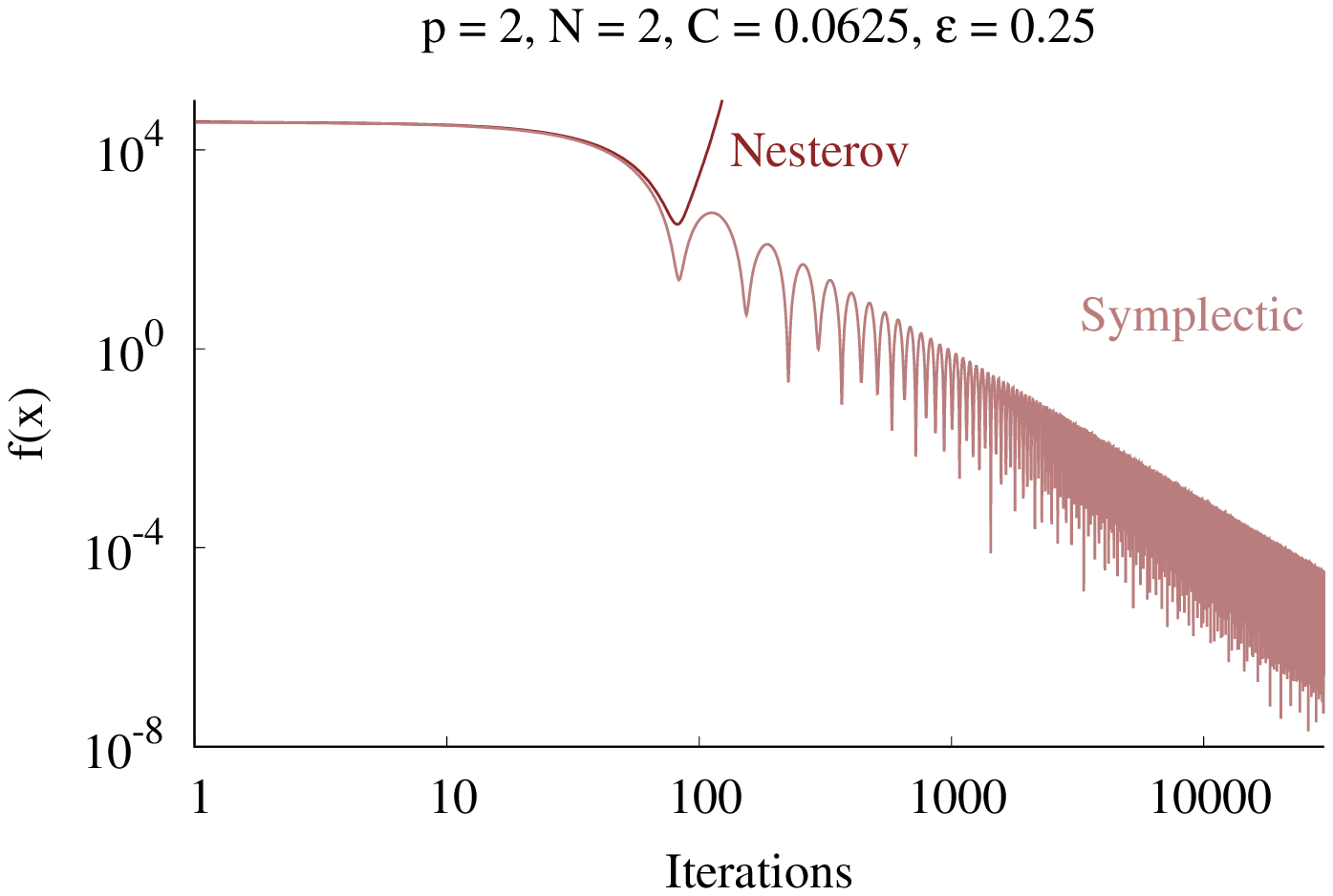} }
\caption{(a) When appropriately tuned, both symplectic optimization
and the dynamic Nesterov discretization simulate the same latent Bregman
dynamics and hence achieve similar convergence rates, here 
approximately $\mathcal{O}(t^{-2.95})$. (b) The symplectic optimization,
however, requires only half of the computational effort of the three-step
generalized Nesterov discretization. (c) Moreover, the inherent stability of
the symplectic optimization admits larger discretization step sizes
and even higher performance improvements.}
\label{fig:nom_iter}
\end{figure*}

Results of this experiment are shown in Figure \ref{fig:nom_iter}a.
We see that the initial convergence rate obtained by symplectic integration 
and the three-step generalized Nesterov discretization are both roughly
$\mathcal{O}(t^{-2.95})$ for this problem.  This should be no surprise,
given that both approaches are stable discretizations of the same 
underlying Bregman dynamics.  The number of iterations to arrive 
near the optimum is accordingly similar for the two algorithms for large
values of the error criterion.  It is smaller for the Nesterov discretization 
in the case of smaller error values.  We return to this 
phenomenon---the increasing rate of the Nesterov discretization as 
it approaches the optimum---in the following section.  

It is important to emphasize that the number of iterations is not the 
same as wall-clock time.  Indeed, the leapfrog integrator that drives our
implementation of symplectic optimization requires only a single 
gradient evaluation per iteration while the three-step generalized 
Nesterov discretization requires two to achieve stability.  Consequently, 
as shown in Figure \ref{fig:nom_iter}b, once we normalize for 
computational cost the symplectic integrator becomes twice as effective 
for large values of the error criterion.  Whether this improvement 
persists in comparison to two-step Nesterov algorithms is an open question.

It is also important to emphasize that with the leapfrog integrator 
we are able to choose larger step sizes than with the three-step 
Nesterov discretization.  This is due to the inherent stability of 
symplectic integration.  In particular, as shown in Figure 
\ref{fig:nom_iter}c, if we increase the step size to $\epsilon = 0.25$ 
we see that the Nesterov discretization quickly diverges while the leapfrog 
integrator remains stable.  The time to arrive near the optimum
decreases uniformly for the leapfrog integrator for this larger
value of the step size.

\section{Achieving Exponential Convergence with a Gradient Flow}

As we have seen in Figure \ref{fig:nom_iter}, the three-step generalized 
Nesterov discretization exhibits a unique behavior 
once it has become sufficiently close to the minimum.  In that neighborhood 
the dynamical Nesterov discretization transitions into an exponential 
rate of convergence towards the minimum and soon surpasses the 
symplectic optimizer.  Interestingly, we have found that this 
behavior does not persist for a quartic objective function, 
$f(x) = \left<x, x\right>^{2}$, suggesting that it requires 
strong convexity of the neighborhood of the objective.  
Exponential convergence of the generalized Nesterov discretization in 
regions of strong convexity of the objective was considered in 
\citet{WibisonoEtAl:2016}.

Because this phase of exponential convergence does not appear 
in symplectic optimization it cannot be a feature of the Bregman 
dynamics themselves.  Instead it must be a side effect of the 
heuristic discretization of the generalized Nesterov discretization, 
which introduced the auxiliary sequence,
\begin{equation*}
y_{n} = \underset{y \in \X}{\mathrm{argmin}}
\left[ f_{p - 1}(y, ; x_{n}) 
+ \frac{N}{\epsilon^{p} \, p} || y - x_{n} ||^{p} \right],
\end{equation*}
or, for the conditions of Section \ref{sec:experiments},
\begin{equation*}
y_{n + 1} = x_{n + 1} 
- \frac{p \, \epsilon^{p} }{2 N} 
\nabla f(x_{n + 1}).
\end{equation*}

For these conditions this sequence actually simulates a 
gradient flow on the configuration manifold, and 
consequently its addition interweaves the Bregman dynamics 
with a gradient flow.  The exact nature of this interweaving, 
however, seems to ensure that the two evolutions characterize
the dynamic Nesterov discretization in different regimes.  Away 
from the minimum of the objective the Bregman dynamics 
dominate, rapidly pulling the system towards the minimum.  
Asymptotically, however, the dynamics dampen and eventually 
the gradient flow becomes dominant.  

For sufficiently well-behaved objectives the emergence of 
the gradient flow allows admits the exponential convergence 
seen in the quadratic objective of Section~\ref{sec:experiments}.  
The gradient flow not only not only stabilizes the dynamic 
Nesterov discretization, it can also provide for even faster 
convergence near the minimum of the objective!

Although symplectic optimization doesn't need a gradient 
flow for stability, it could possibly benefit from the 
potentially exponential convergence it admits.  Fortunately,
incorporating a gradient flow into a symplectic integrator 
is straightforward---instead of trying to approximate the
evolution operator $\exp\left( \epsilon \vec{H}_{\Xi} \right)$ 
we instead try to approximate 
\begin{equation*}
\exp\left( \epsilon \left( 
\vec{H}_{\Xi} + \vec{X}_{\mathrm{GF}} \right) \right),
\end{equation*}
where
\begin{equation*}
\vec{X}_{\mathrm{GF}} = 
- \frac{p \, \epsilon^{p} }{2 N} 
\nabla f(x)
\frac{ \dd }{ \dd x }
\end{equation*} 
is the gradient field generating the gradient flow.
Provided that we construct an appropriate symmetric
splitting then the resulting integrator will enjoy the 
same global error as a symplectic integrator applied to
the extended Hamiltonian system.

For the leapfrog integrator we constructed in Section 
\ref{sec:leapfrog} all we have to do is add 
$\vec{X}_{\mathrm{GF}}$ to the central operator, replacing 
$\exp( \epsilon (\vec{H}_{B3}) )$ with
$\exp( \epsilon (\vec{H}_{B3} + \vec{H}_{\mathrm{GF}}) )$.
Although this combined evolution operator is technically
nonlinear and requires an implicit solution, here we will 
approximate the evolution with the explicit update,
\begin{equation*}
x_{n + 1} 
=
\frac{p}{t^{p + 1}} r_{n + \frac{1}{2}}
- \frac{p \, \epsilon^{p} }{2 N} 
\nabla f(x_{n - 1}).
\end{equation*}

\begin{figure*}
\centering
\subfigure[]{\includegraphics[width=2.8in]
{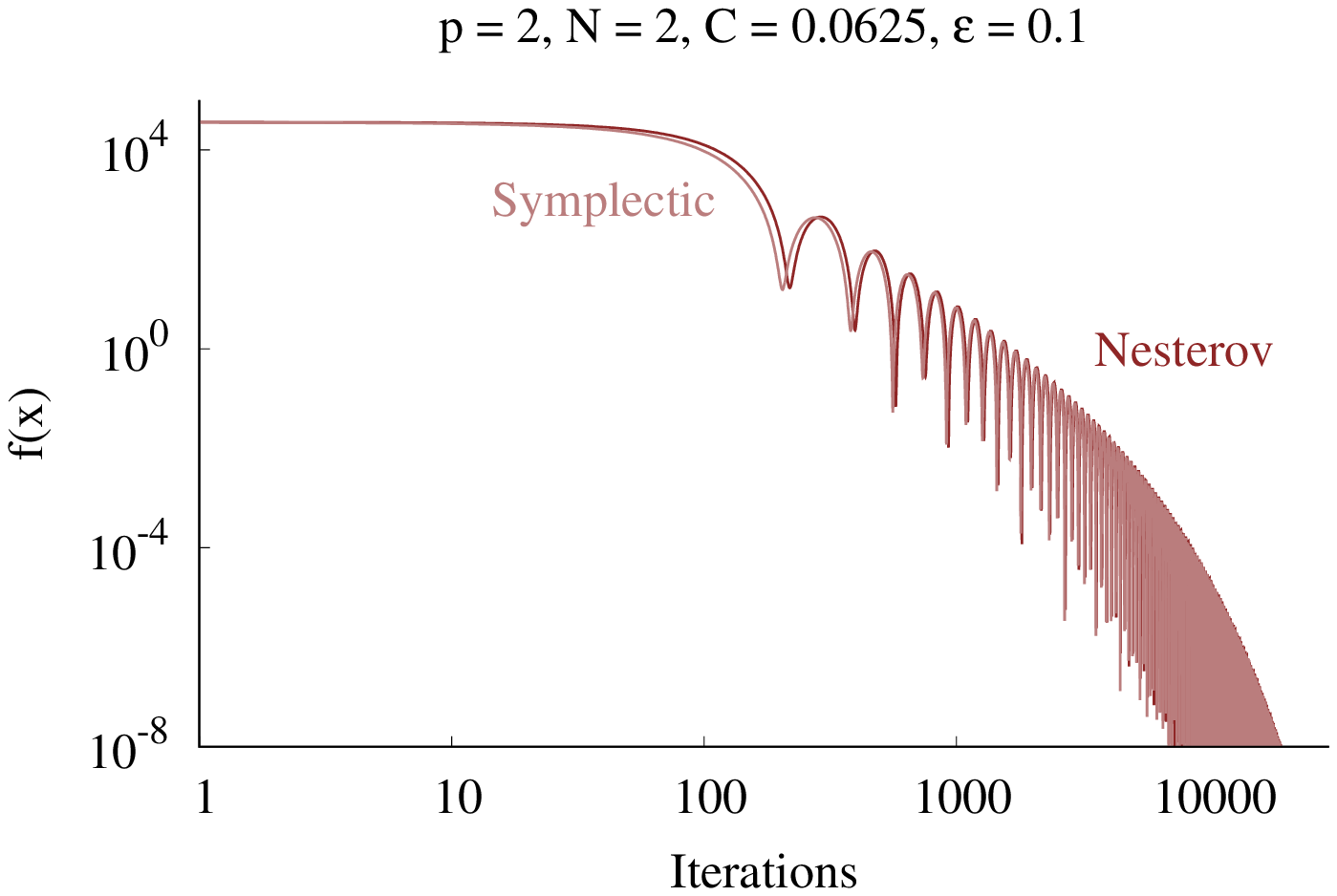}}
\subfigure[]{\includegraphics[width=2.8in]
{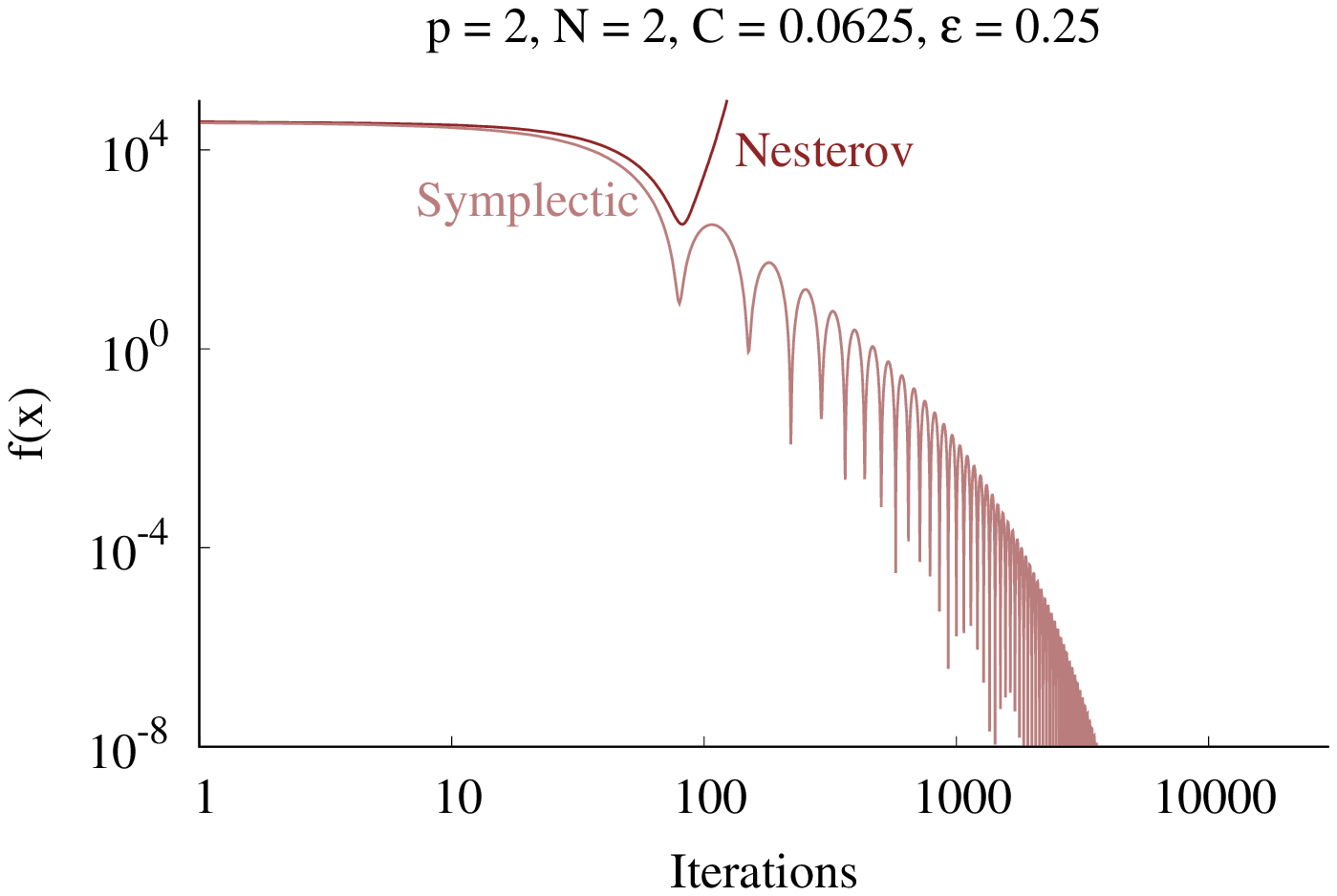}}
\caption{(a) By incorporating gradient flow into the leapfrog 
integration of the Bregman Hamiltonian dynamics we recover 
the same asymptotic exponential convergence near the minimum
of the objective exhibited by the generalized Nesterov discretization. 
(b) These modified Hamiltonian dynamics remain stable even as 
we increase the step size, allowing for more efficient 
computation without compromising the advantageous asymptotic 
behavior.}
\label{fig:with_grad_flow}
\end{figure*}

In the quadratic case where the dynamic Nesterov discretization
exhibited exponential convergence, this modification of
symplectic optimization exhibits the same advantageous
behavior (Figure \ref{fig:with_grad_flow}a).  Moreover,
the modified symplectic optimization maintains its superior 
stability, still allowing for larger step sizes and faster 
practical convergence (Figure \ref{fig:with_grad_flow}b).

Still, while the global error scaling is preserved under 
the addition of the gradient flow and the modified Hamiltonian 
optimization works well empirically, we cannot always expect 
the same stability with the gradient flow.  The problem is
that gradient flow cannot be generated from a Hamiltonian and 
hence the modified discretized evolution cannot preserve the 
symmetries of the underlying dynamics.  In practice we have 
to be careful to tune the gradient flow so that it its 
contributions, including any violations of the dynamical 
symmetries, are negligible until the Hamiltonian dynamics 
have converged close to the minimum of the objective.

\section{Discussion}

\citet{WibisonoEtAl:2016} introduced a dynamical system that 
converged to the minimum of a given objective function at 
the same rate as accelerated Nesterov methods.  Moreover,
by carefully discretizing the Lagrangian representation of 
these dynamics they were able to explicitly derive entire 
families of known accelerated Nesterov discretizations.  Given 
the dynamical system itself, however, discretization is 
more systematically achieved by considering the Hamiltonian 
view of the system and appealing to symplectic integrators. 

In particular, this systematic approach allows us to isolate 
the effects of the dynamics from other modifications, such as 
the gradient flow added to stabilize the original discretization
of the Lagrangian representation of the dynamics.  This 
separation then allows us to analyze the performance of the 
latent Bregman dynamics and that of any amendments independently.

This then positions us to study the general nature 
and optimality of the Bregman dynamics themselves. This study 
may not even be limited to Euclidean configuration spaces
but perhaps also any manifold in a single unified setting.
We discuss details of the systematically geometric construction 
of the Bregman dynamics and possible generalizations in 
Appendix~\ref{apx:geometric}.

This systematic foundation may also allow us to formalize many 
of the empirical behaviors exhibited by Nesterov methods.  
For example, the folk wisdom is that Nesterov methods do not 
perform particularly well when the objective is stochastic.
This behavior, however, is not particularly surprising given
the nature of symplectic integrators.  As discussed in
\citet{Betancourt:2015}, the stochastic variations in the objective 
introduces a bias into symplectic integrators that corrupts
their accuracy by pushing the numerical approximations away 
from the true dynamics.  Intuitively the dynamical evolution 
moves so quickly that the variation in the stochastic objective 
doesn't have sufficient time to average out, unlike slower 
methods such as such as Robbins-Monro that do work well with 
stochastic objectives.  On the other hand, the time dependence
of the Bregman dynamics may provide a way of compensating for 
this bias.  Only with a formal understanding of the Bregman 
dynamics afforded by this new perspective will be able to 
identify the necessary structure.

Unfortunately, the introduction of Hamiltonian symplectic
integrators also complicates the formal analysis of Hamiltonian 
optimization itself.  For example, the accuracy of leapfrog 
integrators comes from cancellations in their symmetric updates, 
but any individual update can have large error.  Hence we 
cannot expect to be able to bound convergence term-by-term.  
Indeed the stability of symplectic integrators is a global
property---the discretized dynamics oscillate around the true 
dynamics and discrete updates will in general deviate away
from the exact dynamics before finally returning.  To understand 
the convergence of the discretized dynamics we instead have 
to take non-local and topological considerations into account,
as is done in \emph{backwards error analysis} 
\citep{McLachlanEtAl:2004, LeimkuhlerEtAl:2004, HairerEtAl:2006}.

Ultimately, however, the direct window into Bregman dynamics 
provided by their Hamiltonian representation and corresponding 
symplectic integration enables not only a better understanding of 
existing accelerated Nesterov methods but also a principled way 
of developing new implementations and generalizations.

\section{Acknowledgements}

We thank Sam Power for helpful comments.

\bibliography{symplectic}
\bibliographystyle{imsart-nameyear}

\setcounter{section}{0}
\renewcommand{\thesection}{\Alph{section}}

\section{Geometric Construction of the Bregman Dynamics} 
\label{apx:geometric}

Symplectic integrators are not only straightforward
to implement and extremely powerful in practice, they
are applicable to Hamiltonian dynamics defined over
any manifold.  This motivates the consideration as to 
whether or not the Bregman dynamics themselves could 
be generalized beyond a Euclidean configuration space
and onto a more general manifold.

In the main paper we heavily utilized the canonical
coordinates of a Euclidean manifold in the construction
of the Bregman Hamiltonian, paying relatively little 
attention to the difference between points in the 
configuration space, $\X$, elements of its local tangent
and cotangent spaces, and even its tangent and cotangent 
bundles.  Here we take a systematic geometric perspective 
in an attempt to highlight some possible paths towards 
generalizing symplectic optimization.

As in the main text we begin with a Euclidean manifold, 
$\X$, equipped with the auxiliary function, $h$.  We
have to be careful, however, as there are two ways of
defining $h$ that lead to the same Bregman dynamics,
and these two approaches may prove to motivate different 
generalizations.

\subsection{Notation}

We will largely follow the notation of \cite{Lee:2013},
although contraction of a vector field with a one-form
will be denoted with $\lrcorner$.

For any manifold, $\X$, we can construct an associated
tangent bundle, $\pi : T\X \rightarrow \X$.  
Correspondingly, any point in the tangent bundle,
$w \in T\X$ identifies both a point in the base manifold,
$\pi(w) \in \X$ and a vector in the local tangent space,
$v(x) \in T_{\pi(w)} \X$.

We can also construct the dual cotangent bundle,
$\varpi : T^{*}X \rightarrow \X$.  Any point in the
cotangent bundle, $z \in T^{*} \X$, identifies both a
point in the base manifold, $\varpi(z) \in \X$ and
a covector in the local cotangent space, 
$p(x) \in T^{*}_{\varpi(z)} \X$.

\subsection{Defining the auxiliary function on 
the configuration space}
\label{sec:geo_config}

The first way to proceed is to define the auxiliary
function directly on the configuration manifold,
\begin{equation*}
h : \X \rightarrow \R.
\end{equation*}
The corresponding Bregman Divergence is a bit ungainly
to write geometrically, requiring liberal use of the
identification of $\X$ with any tangent space, but the
kinetic energy defining the dynamical system has much
cleaner form.  

Given any point in the tangent bundle, $w \in T\X$, we 
can construct a new point in the base manifold by 
translating along the Euclidean connection for some 
time, $T$,
\begin{equation*}
\pi(w) + T \, v(w) = x' \in \X.
\end{equation*}

A natural question is how the auxiliary function, $h$,
changes under this parallel transport,
\begin{equation*}
\Delta h (w, T) = h(\pi(w) + T \, v(w)) - h(\pi(w)),
\end{equation*}
relative to how it would change differentially,
\begin{equation*}
\delta h (w, T) = (\dd h \, \lrcorner \, T \, v(w))(\pi(w)) 
= T \, (\dd h \, \lrcorner \, v(w))(\pi(w)).
\end{equation*}
Setting $T = \exp(- \alpha(t))$ we can then define the
kinetic energy as a function on the time-dependent 
tangent bundle,
\begin{equation*}
K: T\X \times \R \rightarrow \R,
\end{equation*}
where
\begin{align*}
K(w, t) 
&= \Delta h(w, e^{-\alpha(t)}) - \delta h(w, e^{-\alpha(t)})
\\
&=
h(\pi(w) + e^{-\alpha(t)} \, v(w)) - h(\pi(w)) \\
& \quad - e^{-\alpha(t)} (\dd h \, \lrcorner \, v(w))(\pi(w)).
\end{align*}

This kinetic energy allows us to construct the 
time-dependent Bregman Lagrangian, 
\begin{align*}
L 
&: T\X \times \R \rightarrow \R
\\
& (w, t) \mapsto e^{\alpha(t) + \gamma(t)} \left( K(w, t) - U(\pi(w)) \right),
\end{align*}
where, as before,
\begin{align*}
U
:& \X \times \R \rightarrow \R
\\
& (x, t) \mapsto e^{\beta(t)} f(x).
\end{align*}
Taking the natural coordinates for $\X$ this reduces
to exactly the Bregman Lagrangian discussed in the main 
text, hence they are equivalent functions.

We can now build the Bregman Hamiltonian as the Legendre dual 
of $L$ on $T^{*} \X \times \R$. As before we define the components 
of the conjugate momenta as
\begin{equation*}
p^{i} = \frac{ \partial L}{ \partial v_{i}} (x, v, t),
\end{equation*}
with the Hamiltonian given by the Legendre transform
\begin{align*}
H
:& T^{*}\X \times \R \rightarrow \R
\\
& (x, p, t) \mapsto p \left( v (x, p, t) \right) - L(x, v(x, p, t), t).
\end{align*}

In the main text we were able to solve for the velocities 
as a function of the momenta and cleanly substitute then
into the Bregman Hamiltonian with the Legendre transform
of the auxiliary function, $h$.  We have to be careful
here, however, because as $h$ is not defined on the tangent
bundle it does not admit the same Legendre transform as
the Bregman Lagrangian.

Instead we must exploit the Euclidean structure of $\X$.
Because $\X$ is a vector space with dual 
$\mathcal{Y}$, we can perform a Legendre transform from 
$\X$ to $\mathcal{Y}$ to define 
\begin{equation*}
h^{*} : Y \rightarrow \R.
\end{equation*}
On a Euclidean manifold every element of $\mathcal{Y}$ is 
identified with an element of any cotangent space, so this 
conjugate also defines a function
\begin{equation*}
\eta^{*} : T^{*} \X \rightarrow \R.
\end{equation*}
The gradient of $h^{*}$, however, is identified
with a vector field over $\X$.  In particular, the
gradient of $h^{*}$ is equivalent to the Euler vector
field,
\begin{equation*}
E = x^{i} \frac{\partial}{\partial x^{i} }.
\end{equation*}

Utilizing these two objects we can then write the
Bregman Hamiltonian geometrically as
\begin{align*}
H(z, t) 
&= \quad e^{\alpha(t) + \gamma(t) } 
\left( g^{*} (e^{-\gamma(t)}p(z) + \dd h(\varpi(z))) \right.
\\
& \quad
- g^{*} (\dd h(\varpi(z))) - e^{-\gamma(t)} (p(z) \, \lrcorner \, E)(\pi(z))
\\
&\quad
\left. 
+ e^{\beta} f(\pi(z))
\right).
\end{align*}
Once again, in Euclidean coordinates this reduces
to what was presented in the main text.

\subsection{Defining the auxiliary function on 
the tangent bundle}

The other way to proceed is to define the auxiliary
function on the tangent bundle,
\begin{equation*}
h : T\X \rightarrow \R.
\end{equation*}
The corresponding Bregman Divergence is still a bit 
ungainly to write geometrically, but the kinetic energy 
also exhibits a cleaner form.

In this case we exploit the identification of every 
point $x \in \X$ with a vector $u(x) \in T_{x} \X$.  
A point in the tangent bundle, $w$, then defines two
vectors which we can add,
\begin{equation*}
u(\pi(w)) + T \, v(w) = u' \in T_{x} \X.
\end{equation*}
As before we can write
\begin{equation*}
\Delta h (w, T) = h(u(\pi(w)) + T \, v(w)) - h(u(\pi(w))),
\end{equation*}
only now $\Delta h$ is a function on the tangent bundle.

The differential change in $h$, however, is a bit more 
subtle. Because $h$ is now defined as a function on the
tangent bundle, its differential, $\dd h$ is a section 
of $T^{*} T \X$.  In order to contract against the 
differential we need a section of $T T \X$.  Fortunately 
we can use the rigid Euclidean connection of $\X$ to 
define a unique horizontal lift from the vector field 
$v(w)$ over $\X$ to  a vector field $\tilde{v}(w)$ on 
$T\X$.  Using this lift we can write
\begin{equation*}
\delta h (w, T) = (\dd h \, \lrcorner \, T \, \tilde{v}(w))(w) 
= T \, (\dd h \, \lrcorner \, \tilde{v}(w))(w),
\end{equation*}
where again $\delta h$ is now a function on the
tangent bundle.

Putting these together and taking $T = \exp(-\alpha(t))$ we 
get the time-dependent kinetic energy
\begin{equation*}
K: T\X \times \R \rightarrow \R
\end{equation*}
where
\begin{align*}
K(w, t) 
&= \Delta h(w, \exp(-\alpha(t))) - \delta h(w, \exp(-\alpha(t)))
\\
&=
h(u(\pi(w)) + e^{-\alpha(t)} \, v(w)) - h(u(\pi(w))) \\
&- e^{-\alpha(t)} \, (\dd h \, \lrcorner \, \tilde{v}(w))(w).
\end{align*}
Superficially this looks the same as the kinetic
energy derived in Section \ref{sec:geo_config} but each 
term has a subtly different geometric interpretation.  
Still, once we impose Euclidean coordinates we see that 
the function, as well as the corresponding Lagrangian, 
is equivalent.

Continuing to the Hamiltonian, the construction of
a Legendre conjugate for $h$ is simplified as we
simply need to use the same Legendre transform from
the tangent to the cotangent bundle that we use to
map the Lagrangian into a Hamiltonian.  This gives
\begin{equation*}
h^{*} : T^{*} \X \rightarrow \R,
\end{equation*}

The gradient $\dd h^{*}$ is now a section of 
$T^{*} T^{*} \X$ which contracts against sections of 
$T T^{*} \X$.  In order to construct such a section 
we can utilize rigid Euclidean structure of $\X$ once
again.  On the cotangent bundle we have the canonical 
one-form $\theta$ which we can raise to a vector 
field with a musical isomorphism, $\theta^{\sharp}(z)$.  
We can then define a unique horizontal lift of this 
vector field to the vector field 
$\widetilde{\theta}^{\sharp}(z)$ on $T T^{*} \X$ using 
the Euclidean connection.

Similarly, for any element, $z \in T^{*} \X$, of the 
cotangent bundle we can identify an element of the 
tangent bundle by applying a musical isomorphism to 
the canonical one-form to give 
$\theta_{\flat}(z) \in T \X$.

With these objects we can now write the Bregman 
Hamiltonian in the alternative form
\begin{align*}
H(z, t) 
&= \quad
e^{\alpha(t) + \gamma(t) } \left(
h^{*} (e^{-\gamma(t)} p(z) + \dd h(\theta_{\flat}(z))) \right.
\\
& \quad
- h^{*} (\dd h(\theta_{\flat}(z))) 
- e^{- \gamma(t)} (\dd h^{*} \, \lrcorner \, \widetilde{\theta}^{\sharp})(z)
\\
& \quad
\left.
+ e^{\beta} f(\pi(z))
\right).
\end{align*}
Once again, we have a Hamiltonian whose terms have 
a subtly different geometric interpretation but yield 
an equivalent Hamiltonian function on the cotangent bundle.

As we're utilizing the same Euclidean structure of $\X$ 
these two approaches shouldn't give contradictory answers, 
and we see here that they don't.  The two paths, however, 
may illuminate different strategies for generalizing the 
construction of a Bregman dynamical system beyond the 
Euclidean case.  For example, the second approach seems 
particularly appropriate for Riemannian manifolds equipped 
with an appropriate connection.

Regardless of how we might generalize or modify the Bregman
dynamics, provided we maintain a geometric construction the 
discretization will proceed as smoothly as in the Euclidean 
case thanks to the geometric universality of symplectic 
integrators.

\end{document}